      [1999/12/01 v1.4c Il Nuovo Cimento]
\documentclass{cimento}

\usepackage{graphicx}  
\usepackage{xspace}
\usepackage{subfigure}
\usepackage{multirow}

\newcommand{\ttbar}{$t\overline{t}$\xspace}

\newcommand{\Rb}{$R_b$~}

\newcommand{\Vtb}{$V_{tb}$\xspace}

\title{$W$ helicity and constraints on the $Wtb$ vertex at the Tevatron}
\author{C.~Deterre\from{ins:sac}}
\instlist{\inst{ins:sac} CEA/IRFU/SPP - Saclay, France}
\PACSes{\PACSit{14.65.Ha}{Top quarks}\PACSit{12.15.Hh}{Determination of Cabibbo-Kobayashi-Maskawa matrix elements}}
\begin{document}
\maketitle

\begin{abstract}
We present the results obtained by the CDF and D0 collaborations on the extraction of the CKM matrix element \Vtb, 
the $W$ helicity measurements, and the constraints obtained on the $Wtb$ vertex. 

\end{abstract}

\section{Introduction}
In the Standard Model (SM), the top quark is predicted to decay almost exclusively to a $W$ boson and a $b$ quark, and the $Wtb$ coupling is supposed to be purely V-A.
However, physics beyond the SM could be visible in the top decays by changing the branching fraction of the top quark, or by altering the $Wtb$ coupling.
This would be the case for example with a fourth generation of quarks or supersymmetric particles.
A precise measurement of the top and $W$ properties can be used to probe such models of new physics. 
We will first describe the extraction of the Cabibbo-Kobayashi-Maskawa (CKM) matrix element \Vtb.
Then we will present the measurement of the $W$ helicity, and the constraints on anomalous couplings at the $Wtb$ vertex.

\section{Measurement of \Vtb}
The extraction of \Vtb can be done both from the \ttbar decays and the single top production.
In the \ttbar decays, \Vtb can be extracted if the CKM matrix unitarity is assumed.
Since the single top production is directly proportional to \Vtb, it can be extracted directly from the cross-section measurement without such assumptions, but the precision reached is lower because of the small statistics available.

\subsection{Extraction from the \ttbar decays}
Using \ttbar events we can perform a measurement of \Rb which is defined as:
\[
 R_b = \frac{\mathcal{B}(t \to Wb)}{\mathcal{B}(t \to Wq)} = \frac{|V_{tb}|^2}{|V_{td}|^2+|V_{ts}|^2+|V_{tb}|^2} 
\]
where q=d,s,b.
The extraction of \Vtb can be done by assuming $|V_{td}|^2+|V_{ts}|^2+|V_{tb}|^2 = 1$. We then have: $R_b = |V_{tb}|^2$.

The measurement has been performed recently at D0 with a data sample corresponding to an integrated luminosity of 5.4 fb$^{-1}$, 
both in the semileptonic ($\ell$+jets) and dileptonic ($\ell \ell$) channels.
The selected events contain high energetic jets (at least 2 in the $\ell \ell$ channel, at least 3 in the $\ell$+jets channel) and lepton(s) (exactly one for $\ell$+jets and two for the $\ell \ell$ channel).
In the $\ell$+jets channel, an additional cut on the missing transverse energy is applied to reject the background, which is mostly composed of $W$+jets and multijet events.
In the dilepton channel, a topological selection is applied, based on the missing $E_T$ significance of the event, and the scalar sum of the transverse energies of the leptons and leading jet.

The measurement relies on the identification of b-jets, which is done with a neural network algorithm (NN) using the impact parameters of the tracks associated to jets and the properties of secondary vertices.
In the $\ell \ell$ channel, the NN is applied to the two leading jets, and the lowest output of the two is kept to build templates.
In the $\ell$+jets channel, the number of jets passing a cut on the NN output is counted. 
This requirement has an efficiency of 55 $\pm$ 4$\%$ for b-jets and of 1.5 $\pm$ 0.1 $\%$ for light jets.

\begin{figure}[h!]
 \centering
 \includegraphics[width=.87\textwidth]{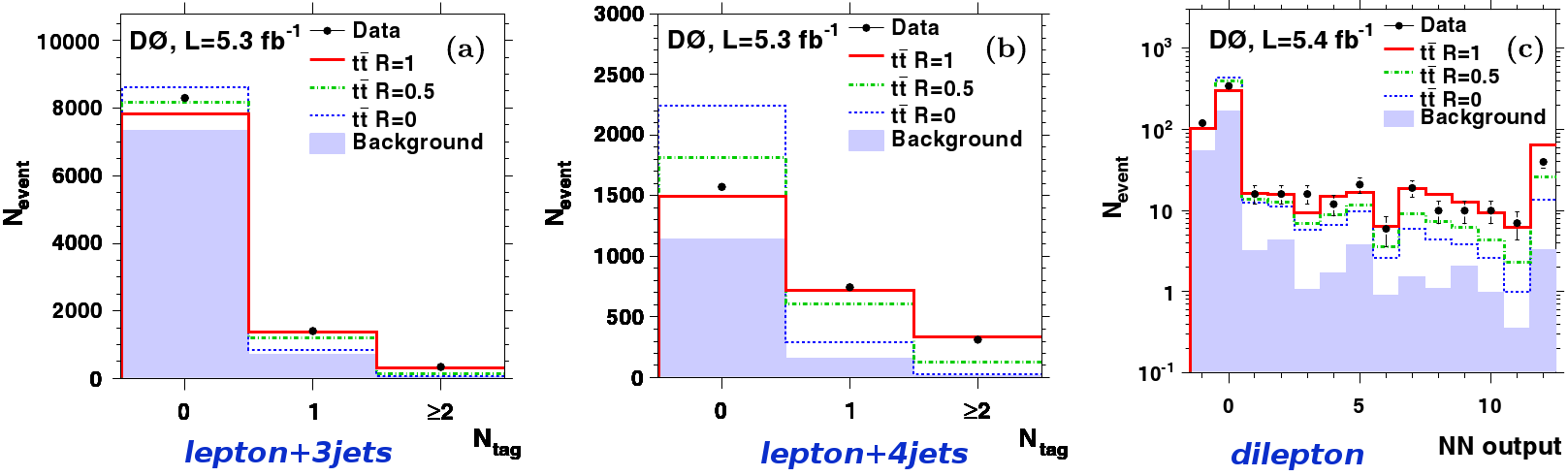}
 \caption{Templates used for the extraction of $R_b$: number of identified b-jets in the $\ell$+3jets (left) and $\ell$+4 jets channels (middle), and output of the b identification algorithm for the $\ell \ell$ channel (right).}
 \label{fig:templates}
\end{figure}

Three decay modes of the \ttbar pair are simulated using Pythia: $t\overline{t} \to bb$, $bq_l$ and $q_lq_l$ where $q_l=d,s$. 
Templates are built for each of these decay channels and the background (see Figure~\ref{fig:templates}), and data is fitted to the templates using a binned likelihood.
A nuisance parameter approach is used to take into account the systematic uncertainties.

The \ttbar production cross-section and $R_b$ were simultaneously extracted with this method. 
D0 obtains~\cite{Rb_D0}: $\sigma_{t\overline{t}} = 7.74 ^{+0.67} _{-0.57}$ (stat+syst) pb and $R_b = 0.90 \pm 0.04$ (stat+syst).
Using the modified frequentist Feldman-Cousins method, the interval in \Vtb at the 95$\%$ confidence level was derived: 0.90~$<|V_{tb}|<$~0.99.

CDF obtained a result using a similar method with 160 pb$^{-1}$ of integrated luminosity~\cite{Rb_CDF}:
$R_b = 1.12 ^{+0.21}_{-0.19}$ (stat) $^{+0.17}_{-0.13}$ (syst), and $|V_{tb}| >$ 0.78 at 95$\%$ C.L.

\subsection{Extraction from the single top cross-section}
This measurement can be performed by assuming the SM single-top cross-sections, a branching fraction of $t \to Wb$ of 1 and the V-A coupling at the $Wtb$ vertex.
The Tevatron combination using results of D0 and CDF with 2.3 to 3.2 fb$^{-1}$ of luminosity yields~\cite{stop_tev}:\\ $|V_{tb}| = 0.88 \pm 0.07$ and $|V_{tb}| > 0.77$ at the 95$\%$ C.L.\\
At D0, a measurement updated with 5.4 fb$^{-1}$ yields~\cite{stop_D0}:\\ $|V_{tb}| = 1.02 ^{+0.10} _{-0.11}$ (stat+syst) and $|V_{tb}| > 0.79$ at the 95$\%$ C.L.

\section{Measurement of the $W$ helicity}

The $W$ boson has three distincts polarizations: it can be longitudinal, left- or right-handed. 
Because of the V-A coupling at the $Wtb$ vertex and the small mass of the $b$ quark compared to the top quark mass, b-quarks from top decays are almost only left-handed.
Since the angular momentum is conserved at the vertex, almost no right-handed $W$ bosons are produced from a top decay.
In the SM, the different helicity fractions are predicted to be: $f_0 = 0.70$ (longitudinal), $f_+ = 0.30$ (right-handed) and $f_- \approx 3.6 \times 10^{-4}$ (left-handed)~\cite{fischer}.
To measure these fractions, two main methods are used: a template method and a matrix element method.

\begin{figure}[h!]
   \begin{minipage}{.45\linewidth}
   \centering
   \includegraphics[width = \textwidth]{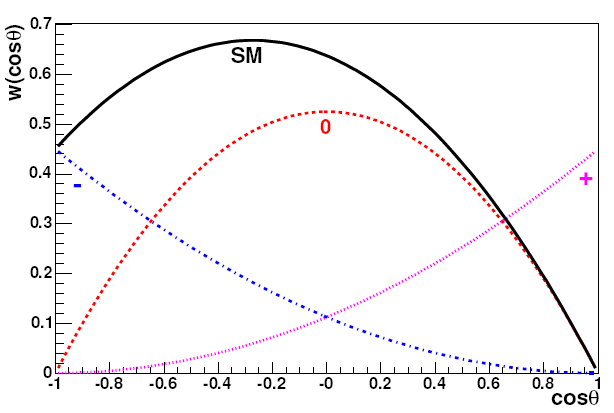}
   \caption{$cos\theta^*$ distributions for different helicities}
   \label{fig:Wtemp}
   \end{minipage}
   \hfill
   \begin{minipage}{.45\linewidth}
   \centering
   \includegraphics[width = \textwidth]{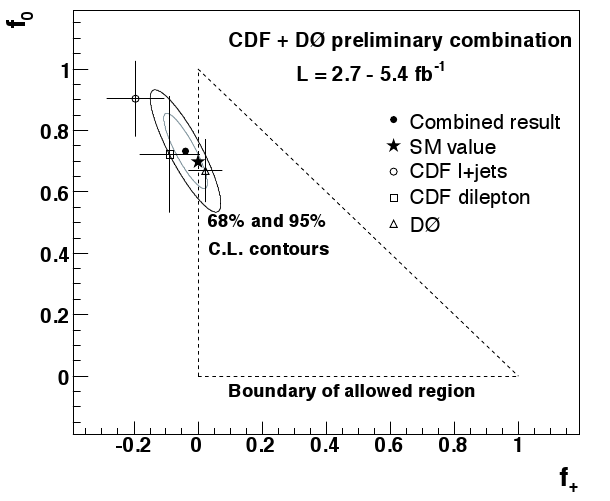}
   \caption{Combination of D0 and CDF results for the $W$ helicity measurement}
   \label{fig:Whel_tev}
   \end{minipage}
  \end{figure}

The template method has been used by CDF in the $\ell \ell$ channel with 5.1 fb$^{-1}$ and by D0 in the $\ell$+jets and $\ell \ell$ channels with 5.4 fb$^{-1}$.
Both measurements use the $cos(\theta^*)$ variable which is very sensitive to the helicity, 
and where $\theta^*$ is defined as the angle between the direction of the incoming top and the charged lepton or down-type quark in the $W$ rest-frame (see Figure~\ref{fig:Wtemp}).
Templates are built for the three different polarizations and the background, taking into account reconstruction effects.

We present here the results obtained with two-dimensional fits of ($f_0$,$f_+$), with the constraint $f_0 + f_+ + f_- = 1$. 
With 4.8 fb$^{-1}$ in the dilepton channel, CDF obtains~\cite{Wheldil_CDF}:\\ 
$f_0 = 0.78 ^{+0.19} _{-0.20}$ (stat) $\pm 0.06$ (syst) and $f_+ = -0.12 ^{+0.11} _{-0.10}$ (stat) $\pm 0.04$ (syst).\\
D0 results in the $\ell$+jets and $\ell \ell$ channels with 5.4 fb$^{-1}$~\cite{Wheldil_D0} are: \\
$f_0 = 0.669 \pm 0.078$ (stat) $\pm 0.065$ (syst) and $f_+ = 0.023 \pm 0.041$ (stat) $\pm 0.034$ (syst).

CDF also used the matrix element method in the $\ell$+jets channel with 2.7 fb$^{-1}$.
The probability for each event to correspond to a certain set of parameters ($f_0$,$f_+$) is computed, by integrating over the PDF, the jet transfer functions, 
and the matrix element which is proportional to the helicity fractions. The fractions obtained are~\cite{Whellj_CDF}: \\
$f_0 = 0.879 \pm 0.106$ (stat) $\pm 0.062$ (syst) and $f_+ = -0.151 \pm 0.067$ (stat) $\pm 0.057$ (syst).

The CDF and D0 measurements have been combined using the BLUE method (best linear unbiased estimator), and the result is~\cite{Whel_tev}:\\
$f_0 = 0.732 \pm 0.063$ (stat) $\pm 0.052$ (syst) and $f_+ = -0.039 \pm 0.034$ (stat) $\pm 0.030$ (syst).\\
Figure~\ref{fig:Whel_tev} shows the individual and combined results with the SM prediction.

%

\section{Constraints on anomalous couplings at the $Wtb$ vertex}

The most general Lagrangian including operators up to dimension five can be written as:
 \[
   \mathcal{L} = -\frac{g}{\sqrt{2}} \overline{b} \gamma^{\mu} V_{tb} (L_V P_L + R_V P_R)tW^-_{\mu} - 
   \frac{g}{\sqrt{2}} \overline{b} \frac{i\sigma^{\mu\nu}q_{\nu} V_{tb}}{M_W}(L_T P_L + R_T P_R)tW^-_{\mu} + h.c. \nonumber
 \]
where $M_W$ and $q_{\nu}$ are the $W$ boson mass and four-momentum,
$L_V$ and $R_V$ are left- and right-handed vector components ($L_V = V_{tb} f_{L_V}$ and $R_V = V_{tb} f_{R_V}$),
and $L_T$ and $R_T$ are left- and right-handed tensor components ($L_T = V_{tb} f_{L_T}$ and $R_T = V_{tb} f_{R_T}$).
The SM predicts:  $L_V = 1$, $R_V = 0$, $L_T = 0$ and $R_T = 0$.
A non-vanishing anomalous coupling would change the rate and kinematics of the single-top production and the $W$ helicity fractions.
In the measurements, two couplings are investigated simultaneously: $L_V$ and one of the anomalous ones, the others being set to zero.

For the single-top, multivariate discriminants (Bayesian neural networks) have been used to distinguish between signals obtained from the different couplings.
Inputs to these discriminants are kinematic variables such as the lepton and jets four-vectors, the missing $E_T$ and b-tagging informations.
D0 obtains limits with 5.4 fb$^{-1}$ shown in Table~\ref{tab:anom}~\cite{anom_D054}.
The combination of single-top results with $W$ helicity measurements can bring an even better discrimination between left- and right-handed couplings.
It has been done at D0 with 2.7 fb$^{-1}$ of integrated luminosity~\cite{anom_D02.7} (see Table~\ref{tab:anom}).

 \begin{table}
  \centering
  \begin{tabular}{ll|lll}
   \multicolumn{2}{c|}{Scenario}                                                      & ($L_V$,$L_T$) & ($L_V$,$R_V$) & ($L_V$,$R_T$) \\
   \hline
   \multirow{2}{*}{Limit on $|V_{tb} f_{X}|^2$} & D0 5.4 fb$^{-1}$~\cite{anom_D054}  & $<$ 0.13      & $<$ 0.93      & $<$ 0.06  \\ 
                                                & D0 2.7 fb$^{-1}$~\cite{anom_D02.7} & $<$ 0.19      & $<$ 0.72      & $<$ 0.20  \\ 

 \end{tabular}
 \caption{Limits on anomalous couplings from the single-top channel (first row) and the combination of single-top and $W$ helicity (second row).}
  \label{tab:anom}
 \end{table}

\section{Conclusion}
Top decays can be used for many searches of physics beyond the Standard Model. 
The CDF and D0 collaborations have been able to put several limits on new physics models. Some of these results are still statistically limited, 
and would benefit from the analysis of the full Tevatron dataset or from a combination with the LHC.
%

\end{document}